\documentclass[pdflatex]{pasj02}
\usepackage[switch,mathlines]{lineno}
\usepackage{url}
\usepackage{mathcomp}
\usepackage{color}
\usepackage{soul}

\Received{$\langle$reception date$\rangle$}
\Accepted{$\langle$acception date$\rangle$}
\Published{$\langle$publication date$\rangle$}

\newcommand{\Kq}{Paper I}
\newcommand{\Kqs}{Paper I }
\newcommand{\Kc}{K25}

\newcommand{\ttiny}{\fontsize{3pt}{3pt}\selectfont }

\graphicspath{{./}{figure/}}

\begin{document}

\title{Mass ratio estimates for overcontact binaries using the derivatives of light curves. II. Systems with deep eclipses}

\author{Shinjirou Kouzuma}

\altaffiltext{}{Faculty of Liberal Arts and Sciences, Chukyo University, 101-2 Yagoto-honmachi, Showa-ku, Nagoya, Aichi 466-8666, Japan}
\email{skouzuma@lets.chukyo-u.ac.jp}


\maketitle

\begin{abstract}
This is the second paper that proposes a simple method for estimating mass ratios using the derivatives of light curves for overcontact binaries. 
In the first paper (Kouzuma 2023), we presented a method to estimate the mass ratios for systems exhibiting a double-peak feature in the second derivatives of their light curves around eclipses. 
This second paper focuses on overcontact systems that are not addressed in the first paper, that is, systems lacking a double-peak in the second derivative. 
A sample of synthetic light curves for overcontact binaries consists of 89,670, covering a parameter space typical of overcontact systems. 
On the basis of a recent study that proposed a new classification scheme using light curve derivatives up to the fourth order, the sample light curves were classified. 
We found that time intervals between two local extrema in the derivatives are associated with the mass ratio in systems exhibiting a high degree of eclipse obscuration. 
Using regression analysis for the identified associations, we derived empirical formulae to estimate the mass ratio and its associated uncertainty. 
The application of our proposed method to real overcontact binary data demonstrated its effectiveness in providing reliable estimates for both values. 
\end{abstract}

%\pagewiselinenumbers

%%%%%%%%%%%%%%%%%%%%%%%%%%%%%%%%%%%%%%%%%%%%%%%%%%%%%%%%%%%%%%%%%%%%%%
\section{Introduction}
This paper presents a simple method for estimating the mass ratios of overcontact eclipsing binaries that were not addressed in \authorcite{Kouzuma2023-ApJ} (\yearcite{Kouzuma2023-ApJ}, hereafter Paper I).
The mass ratio of an overcontact binary system is a fundamental parameter that strongly characterizes the system. 
The most reliable method for determining a mass ratio is to derive a radial velocity curve through repeated spectroscopic observations. 
A more convenient method is to estimate a photometric mass ratio from the light curve (LC) of an eclipsing binary (e.g., \cite{Kopal1959-cbs, Wilson1994-PASP}). 
A photometric mass ratio is generally estimated through LC modeling using an iterative method. 
In this LC analysis, setting appropriate initial parameters is crucial for deriving a reasonable solution, and iterative methods often involve relatively high computational costs. 
In addition, it has been shown that photometric mass ratios can be accurately determined in semi-detached and overcontact systems exhibiting total-annular eclipses; 
in contrast, the derived values may be unreliable when the eclipses are partial \citep{Wilson1994-PASP,Pribulla2003-CoSka,Terrell2005-ApSS}. 

Large photometric surveys have discovered the LCs of numerous overcontact binaries: 
e.g., Kepler \citep{Prsa2011-AJ,Kirk2016-AJ}, 
Lincoln Near-Earth Asteroid Research (LINEAR; \cite{Palaversa2013-AJ}), 
Catalina Real-Time Transient Survey (CRTS; \cite{Drake2014-ApJS}), 
All-Sky Automated Survey for Supernovae (ASAS-SN; \cite{Jayasinghe2018-MNRAS}), 
and Transiting Exoplanet Survey Satellite (TESS; \cite{Prsa2022-ApJS}). 
However, most of them are too faint to obtain radial velocity curves through spectroscopic observations, except when using large telescopes. 
Even if spectroscopic measurements of their radial velocities become easier and more practical in the future, 
obtaining radial velocity curves for more than several thousand binaries would still be highly challenging. 
Therefore, a simplified method to reliably estimate mass ratios from LCs is highly valuable for investigating these binaries, both individually and statistically. 

\Kqs presented an alternative approach for estimating photometric mass ratios of overcontact binaries, introducing a new perspective that uses the derivatives of their LCs. 
This approach requires only the time interval between two local extrema found in the third derivative of a LC. 
Because it requires no iterative procedure, its computational cost is significantly lower. 
Moreover, it provides mass ratio estimates with reasonably reliable uncertainties. 
Several studies have applied the method of \Kqs to real overcontact binaries and have compared the resulting mass ratios with values obtained from iterative approaches such as q-search and Markov Chain Monte Carlo \citep{Sarvari2024-RAA, Baudart2024-RAA, Poro2024-AJ, Poro2025-MNRAS}. 
A total of 23 systems have been examined by the cited studies, and for most of them, the mass ratios agree with those obtained using the other approaches, within the uncertainties estimated by the method of \Kq. 
However, the method presented in \Kqs is applicable only to overcontact binaries whose LCs exhibit a double-peak feature in their second derivatives around the time of an eclipse. 

\authorcite{Kouzuma2025-PASJ} (\yearcite{Kouzuma2025-PASJ}, hereafter \Kc) proposed a new classification scheme for the LCs of overcontact eclipsing binaries. 
K25 demonstrated that the double-peak feature, mentioned above, in the second derivative of a LC is indicative of a total-annular eclipse. 
It has been demonstrated that photometric mass ratios are accurately determined when overcontact binaries exhibit total-annular eclipses (e.g., \cite{Pribulla2003-CoSka, Terrell2005-ApSS}). 
These studies support the validity of the method in \Kqs for effectively estimating mass ratios. 
Moreover, in K25's classification, certain types of overcontact systems that produce LCs lacking the double-peak feature can still exhibit total-annular eclipses or high eclipse obscuration. 
The mass ratios of such systems could potentially be estimated using a method similar to that proposed in \Kq. 
Extending the applicability of mass ratio estimation to a wider variety of LCs enables more efficient use of archival data. 
It also enhances statistical analyses of overcontact binaries and supports large-scale, spectroscopy-independent studies.

This study focuses on four types of overcontact binaries proposed in \Kc, excluding the type whose LCs exhibit the double-peak feature in their second derivatives. 
Using the methodology in \Kq, we develop new approaches to estimate mass ratios for two of the four types. 
Section \ref{Sec_D} introduces a sample of synthetic LCs for overcontact systems. 
Section \ref{Sec_M} describes methods to estimate mass ratios of overcontact binaries. 
In section \ref{Sec_app}, we introduce real binary data to examine the effectiveness of our proposed methods. 
Section \ref{Sec_Result} presents its results and discusses their effectiveness. 
The summary of this study is provided in section \ref{Sec_Summary}.

%%%%%%%%%%%%%%%%%%%%%%%%%%%%%%%%%%%%%%%%%%%%%%%%%%%%%%%%%%%%%%%%%%%%%%
\section{Data}\label{Sec_D}
In this study, we used the synthetic LC dataset for overcontact binaries from \Kc, which was generated with the PHOEBE 2.4 code \citep{Prsa2016-ApJS, Conroy2020-ApJS}. 
This dataset consists of 89670 LCs computed across a wide range of overcontact binary parameters, including the mass ratio ($q = M_\mathrm{s}/M_\mathrm{p} = 0.05$--$0.95$, in steps of 0.1), orbital inclination ($i = 30\tcdegree$--$90\tcdegree$, $1\tcdegree$), fill-out factor ($f = 0.2$--$0.8$, $0.3$), and stellar temperatures ($T_\mathrm{p}$ and $T_\mathrm{s} = 4000$--$10000$ K, $1000$ K). 
The indices `p' and `s' refer to the primary and secondary, respectively; the primary star is defined as the more massive component of the binary. 

Several predictions have been reported for the minimum mass ratio of overcontact binaries, in which the ratio is likely to fall within the range of 0.044--0.09 (e.g., \cite{Rasio1995-ApJ,Li2006-MNRAS,Yang2015-AJ}). 
On the basis of these studies, this work adopted a lower limit of 0.05 for the mass ratio. 

Each synthetic LC was binned into 100 phase intervals. 
We also computed numerical derivatives of each LC with respect to time, up to the fourth order.

\begin{figure*}[ht]
 \begin{center}
  \includegraphics[width=0.48\textwidth]{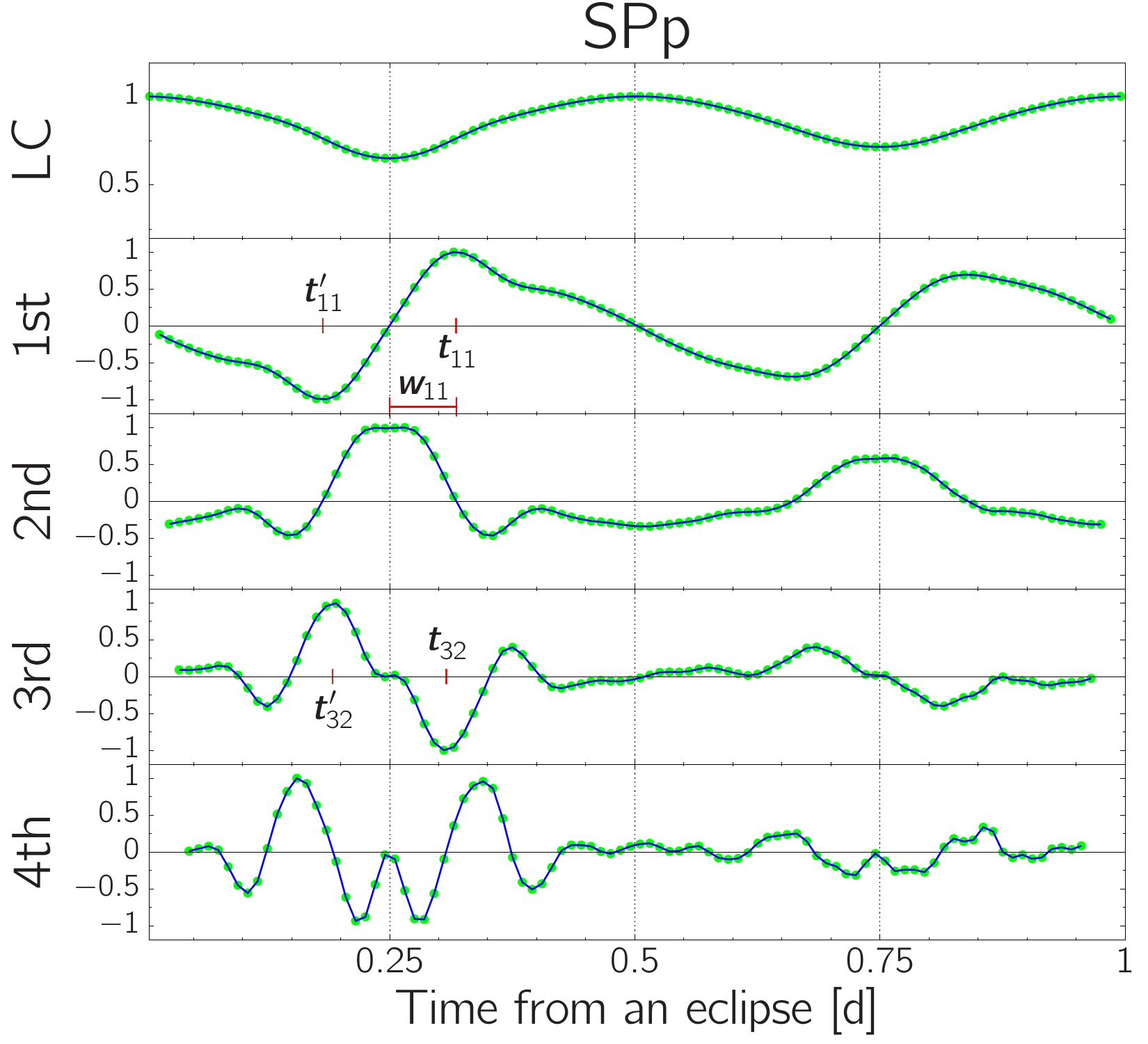}
  \includegraphics[width=0.48\textwidth]{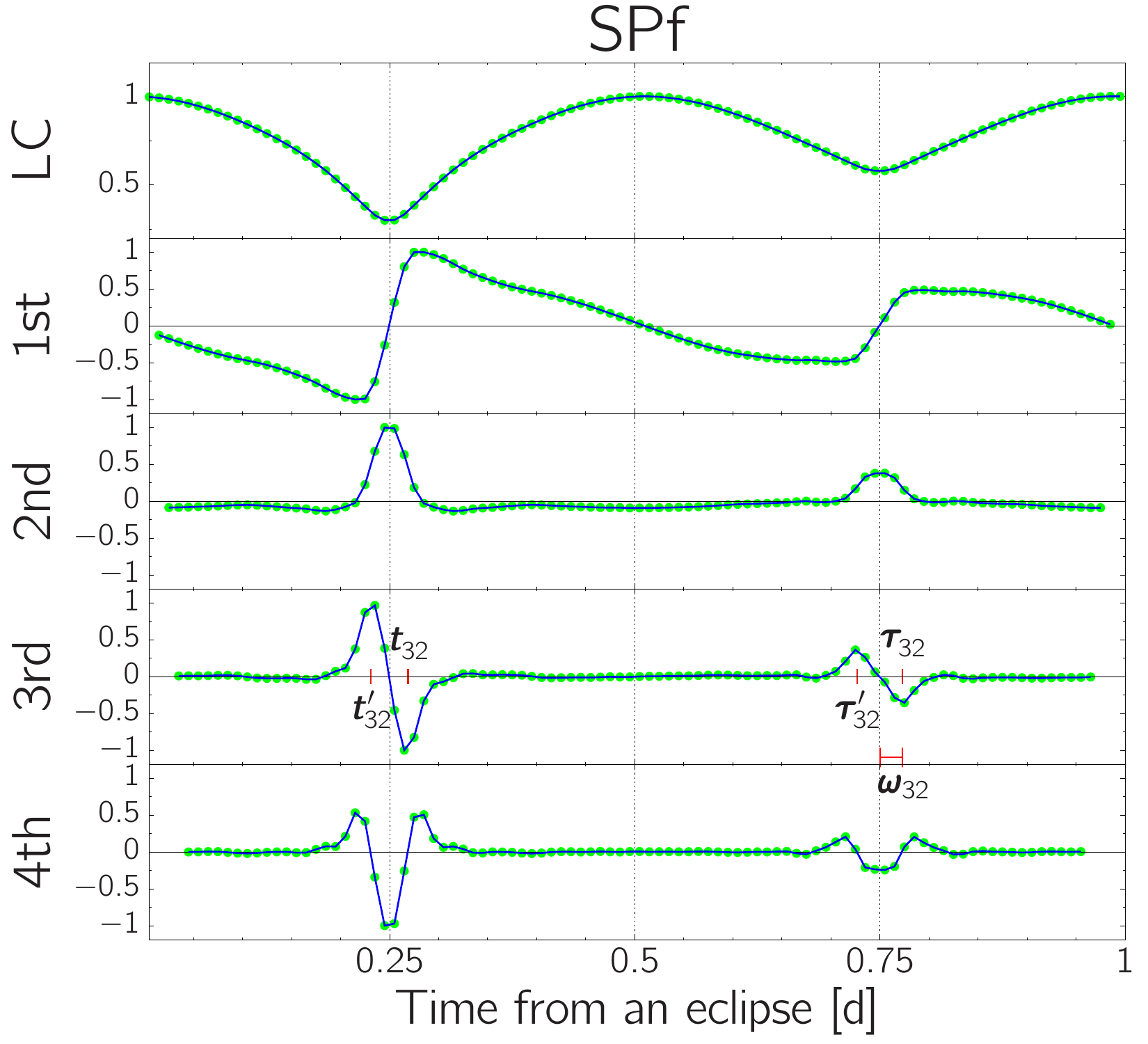}
 \end{center}
 \caption{Light curves and their first through fourth derivatives (from top to bottom) for representative sample binaries of the SPp and SPf types. 
			Each derivative is normalized to unity. 
			Key times used for estimating mass ratios and their associated uncertainties are labeled. 
		 {Alt text: Two panels illustrate example light curves and their derivatives for the SPp and SPf types. 
					The x-axes show the time from an eclipse, ranging from 0 to 1 day. }
 \label{fig_LCs}}
\end{figure*}
%%%%%%%%%%%%%%%%%%%%%%%%%%%%%%%%%%%%%%%%%%%%%%%%%%%%%%%%%%%%%%%%%%%%%%
\section{Method}\label{Sec_M}
Our method for estimating mass ratios is based on a relationship between the mass ratio and the time interval between local extrema found in the derivatives of a LC. 
In this section, we present the key time intervals and their relationships with the mass ratio. 
We also provide uncertainty values for the estimated mass ratios. 

%%%%%%%%%%%%%%%%%%%%%%%%%%%%%%%%%%%%%%%%%%%%%%%%%
\subsection{Finding key values}
We first classified the synthesized LCs according to the classification method introduced by \Kc: DP, SPp, SPb, SPf, and SPs-types. 
DP-type LCs, which were already addressed in \Kq, are excluded from this study. 
For each of the remaining four types, we thoroughly examined the associations between the mass ratios and all possible time intervals derived from two local extrema in the derivatives of LCs. 
As a result, for SPp- and SPf-type LCs, several time-intervals were found to be associated with the mass ratios. 
By comparing the accuracies of the mass ratios estimated from each relationship, we finally determined that the following two $W$ values are the most appropriate: 
\begin{align}
	W_{\text{SPp}} &= \frac{P}{t_{32} - t'_{11}} = \frac{P}{w_{32} + w'_{11}}, \label{Eq_Wvalue-SPp} \\
	W_{\text{SPf}} &= \frac{P}{\tau'_{32} - t_{32}} = \frac{P}{0.5P - (\omega'_{32} + w_{32})}, \label{Eq_Wvalue-SPf} 
\end{align}
for SPp and SPf systems, respectively. 
Here, $P$ denotes the orbital period. 
The symbols $t_{ij}$ and $\tau_{ij}$ represent the times at either a local maximum or minimum around the primary (deeper) and secondary (shallower) minima, respectively. 
Additionally, we define $w_{ij}$ and $\omega_{ij}$ as the time intervals between $t_{ij}$ or $\tau_{ij}$ and the corresponding primary or secondary eclipse times, respectively. 

We found that the mass ratios of the SPp systems also have strong associations with the time interval used in \Kqs (i.e., $t_{32}-t'_{32}$). 
However, we ultimately determined that the proposed interval in equation (\ref{Eq_Wvalue-SPp}) can yield more accurate estimates of the mass ratio. 
 
Figure \ref{fig_LCs} shows examples of LCs and their derivatives for SPp and SPf systems. 
We used 2991 SPp and 4269 SPf systems. 
Note that both types are overcontact systems with a high eclipse obscuration (see \Kc). 
In both panels of figure \ref{fig_LCs}, the times used to compute the $W$ values in equations (\ref{Eq_Wvalue-SPp}) and (\ref{Eq_Wvalue-SPf}) are labeled.

\begin{figure*}[]
 \begin{center}
  \includegraphics[width=0.48\textwidth]{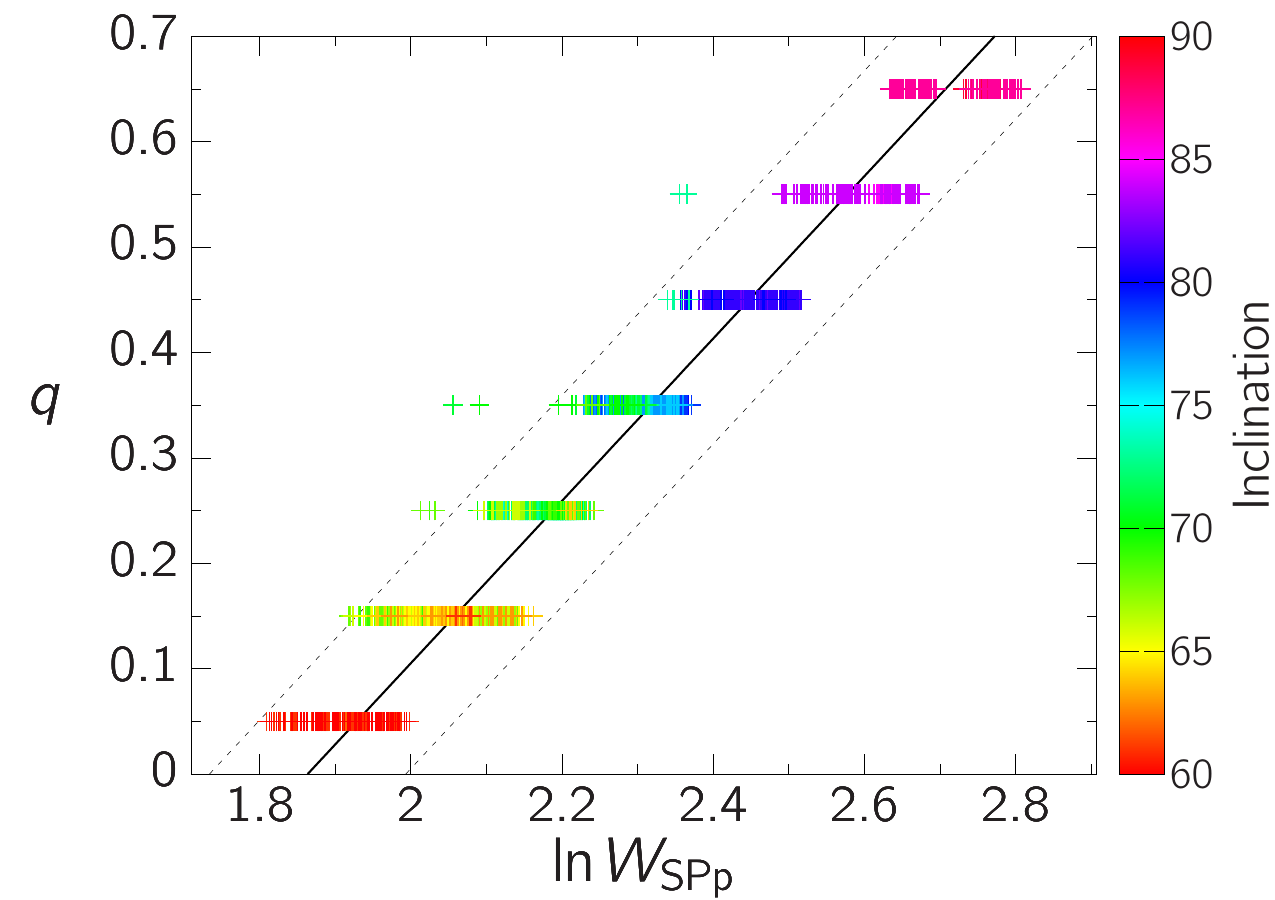}
  \includegraphics[width=0.48\textwidth]{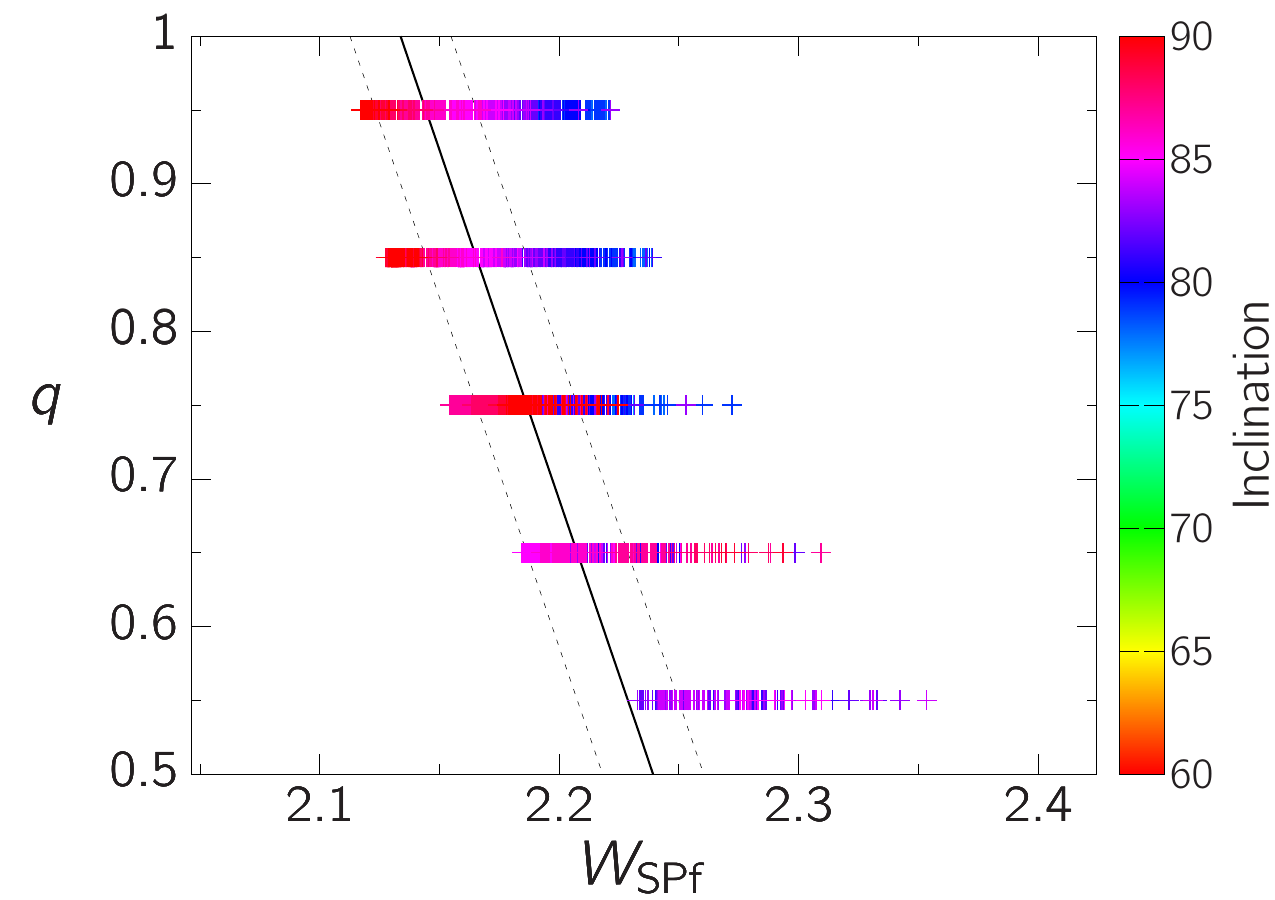}
 \end{center}
 \caption{Relationships between $W$ (equations \ref{Eq_Wvalue-SPp} or \ref{Eq_Wvalue-SPf}) and the mass ratio $q$. 
		  The solid lines represent the regression lines, accompanied by pairs of parallel dashed lines that indicate, for reference, mass ratios 0.1 higher and lower than the regression lines. 
		  The color gradient represents the orbital inclination angle, ranging from $60\tcdegree$ (orange) to $90\tcdegree$ (red). 
		 {Alt text: Two scatter plots for the synthesized SPp-type (left) and SPf-type (right) LCs. 
					The x-axes show $W$ values, ln-scaled, ranging from 1.8 to 2.8 for the SPp systems and from 2.0 to 2.4 for the SPf systems. 
					The y-axes show the mass ratio, ranging from 0 to 0.7 for the SPp systems and from 0.5 to 1.0 for SPf systems. }
 \label{fig_W-q}}
\end{figure*}
%%%%%%%%%%%%%%%%%%%%%%%%%%%%%%%%%%%%%%%%%%%%%%%%%
\subsection{Mass ratio estimation}\label{Sec_Estimation}
Figure \ref{fig_W-q} demonstrates relationships between the $W$ values and mass ratio. 
The solid lines represent regression lines which were fitted to minimize the sum of squares of the residuals in $W$. 
These regression lines are expressed as follows: 
\begin{align}
	q_\text{\tiny SPp} &= 0.770 \ln W_{\text{SPp} - 1.435}, \label{q-estimator-SPp} \\
	q_\text{\tiny SPf} &= -4.739 W_{\text{SPf}} + 11.112, \label{q-estimator-SPf} 
\end{align}
for the SPp and SPf systems, respectively. 
These equations enable the estimation of photometric mass ratios for SPp- and SPf-type LCs based on the key values $W_\text{SPp}$ and $W_\text{SPf}$. 

Pairs of parallel dashed lines for reference, in figure \ref{fig_W-q}, represent mass ratios of higher and lower by 0.1 than the regression lines. 
For the DP systems addressed in \Kq, 99\% of them were within the parallel lines. 
The $W$ values for 99.2\% of the SPp systems and 50.9\% of the SPf systems fell within the range of $q \pm 0.1$. 

In figure \ref{fig_W-q}, a dispersion is observed in both the $W_\text{SPp}$ and $W_\text{SPf}$ values for a given mass ratio. 
The primary factor for the dispersion is the difference in inclination, as also mentioned in \Kq. 
In particular, the SPf systems display a relatively larger dispersion in $W$ values, which results in lower accuracy in our mass-ratio estimation. 
The standard deviations of the residuals along $W$ from the regression lines are $\sigma_{W_\text{SPp}}=$ 0.046 and $\sigma_{W_\text{SPf}}=0.027$, corresponding to those for the mass ratio of $\sigma_{q_{\text{\ttiny SPp}}}=$ 0.035 and $\sigma_{q_\text{\ttiny SPf}}=0.128$, respectively. 
This dispersion contributes to the uncertainty of the estimated mass ratio in this method.

%%%%%%%%%%%%%%%%%%%%%%%%%%%%%%%%%%%%%%%%%%%%%%%%%
\subsection{Uncertainty}
The uncertainties in the mass ratios estimated with our method can be derived in the same manner as in \Kq, which considers two types of uncertainties: 
the measurement uncertainty in the timings of local extrema ($\delta W$), and the uncertainty in the empirical formulae ($\sigma_W$; equations \ref{q-estimator-SPp} and \ref{q-estimator-SPf}). 
The uncertainty of the orbital period $P$ is assumed to be sufficiently small. 

Under the assumption that the two local extrema in the derivatives of a LC are perfectly symmetric (i.e., $w_{11}=w'_{11}$, $w_{32}=w'_{32}$, and $\omega_{32}=\omega'_{32}$), the standard uncertainties of $W_\text{SPp}$ and $W_\text{SPf}$ are
\begin{align}
	\delta W_\text{SPp} &= \frac{W_\text{SPp}^2}{P} \sqrt{\frac{(w_{32}-w'_{32})^2 + (w_{11}-w'_{11})^2}{2}}, \\
	\delta W_\text{SPf} &= \frac{W_\text{SPf}^2}{P} \sqrt{\frac{(\omega_{32}-\omega'_{32})^2 + (w_{32}-w'_{32})^2}{2}}, 
\end{align}
respectively. 

On the basis of equations (\ref{q-estimator-SPp}) and (\ref{q-estimator-SPf}) and the standard deviations in section \ref{Sec_Estimation}, the uncertainty of a mass ratio estimated using our proposed method is derived as follows: 
\begin{align}
	\delta q_\text{\tiny SPp} &= 0.770 \sqrt{\sigma_{W_\text{\tiny SPp}}^2 + \left(\frac{\delta W_\text{\tiny SPp}}{W_\text{\tiny SPp}} \right)^2} \\
							  &\simeq \sqrt{0.035^2 + \left(0.770 \frac{\delta W_\text{\tiny SPp}}{W_\text{\tiny SPp}} \right)^2}, \\
	\delta q_\text{\tiny SPf} &= 4.739 \sqrt{\sigma_{W_\text{\tiny SPf}}^2 + \delta W_\text{\tiny SPf}^2} \\
							  &\simeq \sqrt{0.128^2 + \left(4.739 \delta W_\text{\tiny SPf}\right)^2}. 
\end{align}
An expanded uncertainty is obtained through the multiplication of these values by an appropriate coverage factor.

%%%%%%%%%%%%%%%%%%%%%%%%%%%%%%%%%%%%%%%%%%%%%%%%%%%%%%%%%%%%%%%%%%%%%%
\section{Application to real binary data}\label{Sec_app}
To evaluate the effectiveness of the proposed method, we apply it to real overcontact binaries with spectroscopically determined mass ratios. 
In the catalog by \citet{Latkovic2021-ApJS}, 159 overcontact systems have entries of spectroscopic mass ratios. 
We extracted their LCs from the TESS and Kepler archival data, and LCs for 153 systems were derived. 
Each LC was phase-folded using the orbital period determined via periodogram analysis. 
The phase was divided into 100 bins, and the data were averaged within each bin. 
We also derived numerical derivatives up to the fourth order from the binned LCs. 
When two or more LCs were available for a binary, the better-quality one was selected in such a manner that its derivatives were less noisy and smoother than the other one. 
Apparent outliers in the LCs were removed in advance. 

We classified the selected LCs with the method proposed by K25 and identified 13 SPp and 6 SPf systems. 
These SPp (SPf) systems are classified as 8 (2) W- and 5 (4) A-type systems in Latkovi{\'c}'s catalog. 
These two categories are subtypes of W Ursae Majoris (W UMa) binaries introduced by \citet{Binnendijk1970-VA}.
We then applied our proposed method to these LCs and estimated their mass ratios together with their uncertainties. 
We collected spectroscopic mass ratios and their uncertainties from the original publications, where they were derived from radial velocity curves. 

For reference, we also compare the photometric mass ratios reported in the literature with those estimated by the proposed method. 
Sample systems were extracted from Latkovi{\'c}'s catalog, and their LCs were derived from the TESS and Kepler archival data. 
Following the same procedure as above, we identified 16 SPp (7 W-type and 9 A-type) and 20 SPf systems (13 W-type and 7 A-type). 
The proposed method was also applied to these systems. 
Note that none of these systems have spectroscopic mass ratios available in the catalog.

\begin{figure}[]
 \begin{center}
  \includegraphics[width=0.48\textwidth]{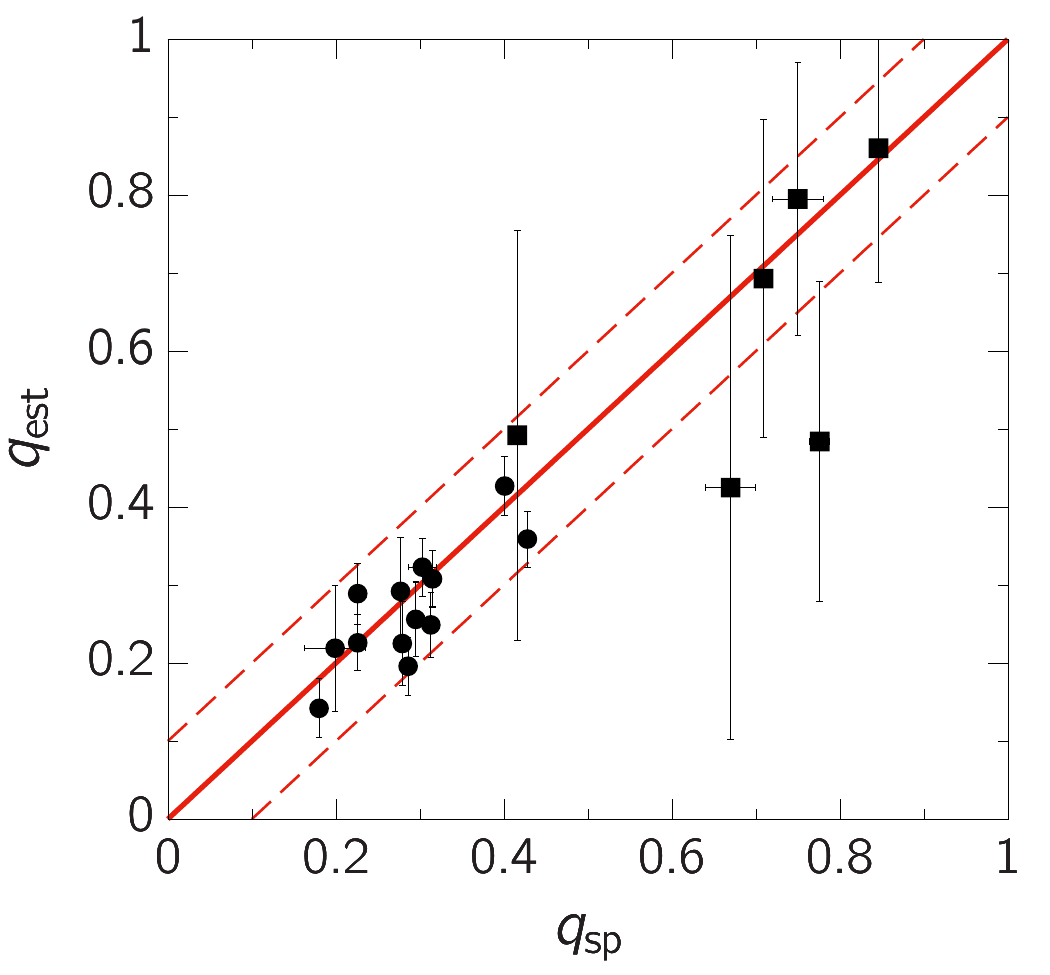}
 \end{center}
 \caption{Comparison between our estimated ($q_\text{est}$) and the literature's spectroscopic ($q_\text{sp}$) mass ratios. 
		Filled circles and squares represent SPp and SPf systems, respectively. 
		  The red solid and dashed lines show $q_\text{est} = q_\text{sp}$ and $q_\text{est} = q_\text{sp} \pm 0.1$, respectively. 
		 {Alt text: Scatter plot of our estimated and literature's spectroscopic mass ratios. 
					The x- and y-axes show the spectroscopic and estimated mass ratios, respectively. 
					Both axes range from 0 to 1. }
 \label{fig:qsp-qest}}
\end{figure}
\begin{figure}[]
 \begin{center}
  \includegraphics[width=0.48\textwidth]{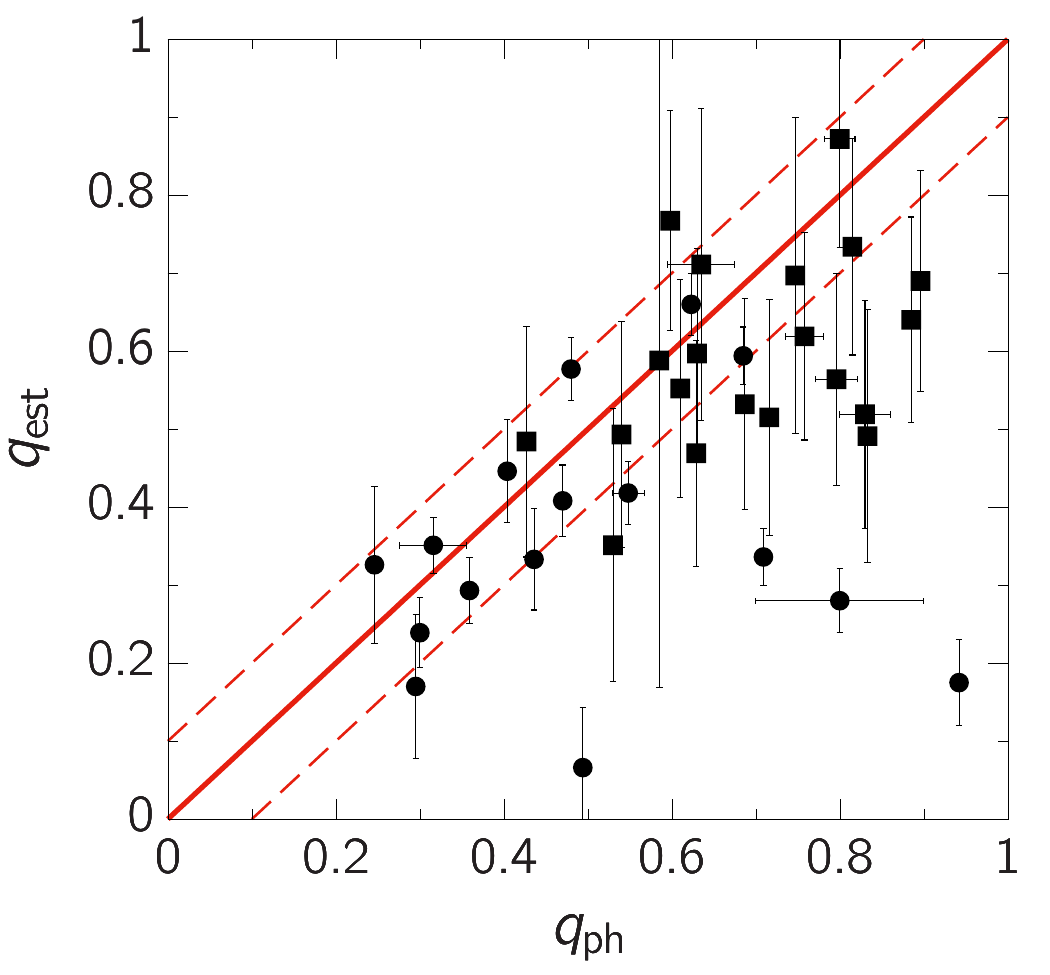}
 \end{center}
 \caption{Comparison between our estimated ($q_\text{est}$) and the literature's photometric ($q_\text{ph}$) mass ratios. 
		Symbols are the same as in figure \ref{fig:qsp-qest}. 
		 {Alt text: Scatter plot of our estimated and literature's photometric mass ratios. 
					The x- and y-axes show the photometric and estimated mass ratios, respectively. 
					Both axes range from 0 to 1. }
 \label{fig:qph-qest}}
\end{figure}
%%%%%%%%%%%%%%%%%%%%%%%%%%%%%%%%%%%%%%%%%%%%%%%%%%%%%%%%%%%%%%%%%%%%%%
\section{Results and Discussion}\label{Sec_Result}
\subsection{Comparison with spectroscopic mass ratios}
Figure \ref{fig:qsp-qest} shows a comparison between our mass-ratio estimates ($q_\text{est}$) and corresponding spectroscopic mass ratios ($q_\text{sp}$). 
The standard deviations of ($q_\text{est}-q_\text{sp}$) are 0.046 and 0.158 for the SPp and SPf systems, respectively. 
The value for the SPf systems is slightly larger than that obtained from the synthesized LCs. 
This may be due to the small sample size. 
Additionally, it could be attributed to observational errors, as also noted in \Kq. 
This discrepancy indicates that the accuracy of our mass-ratio estimation depends on the quality of the observed LCs, particularly their noise level and time resolution. 

The estimated mass ratios for 69\% of the SPp systems and 83\% of the SPf systems are consistent with their spectroscopic values. 
The percentage for the SPf systems exceeds the expected value (i.e., 68\%) based on the standard uncertainty. 
Furthermore, the estimated mass ratios for 100\% of the SPp systems and 67\% of the SPf systems fall within the range of $q_\text{sp} \pm 0.1$. 
Similarly, the percentage for the SPf systems is clearly larger than that for the synthesized LCs mentioned in section \ref{Sec_Estimation} (i.e., 50.9\%). 
These elevated values for the SPf systems are likely attributable to the small sample size. 
If the sample size is sufficiently large (e.g., a few dozen), both percentages for the SPf systems are likely to converge toward the expected values, as observed for the SPp systems. 
The above results support the validity and practical applicability of the proposed method for estimating mass ratios and their associated uncertainties using SPp- and SPf-type LCs. 

All the times in equations (\ref{Eq_Wvalue-SPp}) and (\ref{Eq_Wvalue-SPf}) are closely associated with those used in \Kqs (i.e., $t'_{32}$ and $t_{32}$). 
In other words, these times correspond to the moments immediately after (for $t'_{11}$ or $\tau'_{32}$) or before (for $t_{32}$) the points at which the rates of change in luminosity reach their minimum or maximum, respectively (see also figure 4 in \Kq). 
In a manner similar to that discussed in \Kq, the time intervals $t_{32}-t'_{11}$ and $\tau'_{32}-t_{32}$ in equations (\ref{Eq_Wvalue-SPp}) and (\ref{Eq_Wvalue-SPf}) are, respectively, decreasing and increasing functions of the relative size of the smaller star. 
Therefore, our proposed method is expected to provide effective estimates of the mass ratios for SPp and SPf overcontact systems.

%%%%%%%%%%%%%%%%%%%%%%%%%%%%%%%%%%%%%%%%%%%%%%%%%
\subsection{Comparison with conventional photometric mass ratios}
We also present, for reference, a comparison between our mass-ratio estimates ($q_\text{est}$) and corresponding photometric mass ratios reported in the literature ($q_\text{ph}$) in figure \ref{fig:qph-qest}. 
Note that photometric mass ratios can generally be estimated accurately for systems exhibiting total-annular eclipses \citep{Wilson1994-PASP,Pribulla2003-CoSka,Terrell2005-ApSS}; 
however, in other systems, the associated uncertainties can be substantial, and the resulting estimates should be treated with caution. 

The estimated mass ratios for 25\% of the SPp systems and 45\% of the SPf systems agree with their reported photometric values. 
Several studies have suggested that the uncertainties of photometric mass ratios are likely to be unrealistically small (e.g., \cite{Maceroni1997-PASP, Rucinski2001-AJ1007, Pavlovski2009-MNRAS}). 
According to the error analysis of the LC solutions performed by \citet{Liu2021-PASP}, the relative errors of photometric mass ratios are generally between 10\% and 20\% for partially eclipsing overcontact binaries. 
In our case, if the uncertainties of $q_\text{ph}$ are assumed to be approximately 15\% of their values, then $\sim 68$\% of the estimated mass ratios for both samples agree with the photometric mass ratios. 

In figure \ref{fig:qph-qest}, although the two mass ratios generally agree, several systems show significant discrepancies in $|q_\text{est}-q_\text{ph}|$ value. 
On the basis of a survey of the cited and related literature, we found that the discrepancies likely arise from the analysis method. 
Most of the photometric mass ratios were determined using the q-search method. 
In the q-search method, photometric mass ratios are determined by identifying the minimum in a plot of the sum of squared residuals versus the trial mass ratio. 
Most systems with significant discrepancies exhibited broad and relatively flat minima in the plot. 
In this situation, the determined mass ratios should have large uncertainties and thus be unreliable \citep{Zhang2017-MNRAS}. 
Indeed, in some cases, the mass ratios obtained in other studies differed substantially from the values listed in Latkovi{\'c}'s catalog.

%%%%%%%%%%%%%%%%%%%%%%%%%%%%%%%%%%%%%%%%%%%%%%%%%
\subsection{Influence of starspots}
Starspots can affect the shape of a LC. 
However, if the brightness of the starspots does not vary significantly over a few days, their impact is expected to be minimal. 
This is because our method uses the derivatives of the LC, and the derivative of any constant light contribution would be zero, effectively removing its influence. 
\Kqs demonstrated this by applying the method to synthetic LCs both with and without a constant third light, showing that the mass ratio estimate remains unaffected. 
The same reasoning should apply to starspots, when their brightness is assumed to be nearly constant. 

\citet{Poro2024-AJ} actually examined the impact of a starspot on the method of \Kq, by comparing the mass ratios estimated for 21 systems with and without starspots. 
They found that the difference in the estimated mass ratios was less than 2\% in each instance. 
Consequently, starspots are expected to have little effect on our mass ratio estimates using the derivatives of LCs, at least as long as the brightness of the starspots does not vary significantly over time.

%%%%%%%%%%%%%%%%%%%%%%%%%%%%%%%%%%%%%%%%%%%%%%%%%%%%%%%%%%%%%%%%%%%%%%
\section{Summary}\label{Sec_Summary}
We have proposed a new method for estimating photometric mass ratios from SPp- and SPf-type LCs of overcontact binaries. 
Our proposed method requires derivatives up to the fourth order and the time interval between two local extrema in their derivatives, without requiring an iterative procedure. 
This work extends the methodology presented in \Kqs to other overcontact eclipsing binaries. 
In particular, the present study addressed systems with deep eclipses, whereas \Kqs focused on those with total-annular eclipses. 

The proposed method can estimate the mass ratios of SPp and SPf systems with standard uncertainties of at least 0.035 and 0.128, respectively. 
An application to real binary data demonstrated that the estimated mass ratios for $\sim 69$\% of the SPp systems and $\sim 83\%$ of the SPf systems agreed with their spectroscopic mass ratios within the estimated uncertainties. 
The results in the application indicate that our proposed method effectively estimates the mass ratios and their associated uncertainties. 
While the results for SPf systems are promising, the relatively large uncertainties and limited sample size suggest that further validation using a more extensive dataset would be beneficial.

\begin{ack}
The author would like to thank the anonymous referee for helpful comments and suggestions which improved the paper. 
This work was supported by JSPS KAKENHI Grant Number 25K07358. 
This paper includes data collected by the TESS mission. Funding for the TESS mission is provided by the NASA’s Science Mission Directorate. 
This paper includes data collected by the Kepler mission and obtained from the MAST data archive at the Space Telescope Science Institute (STScI). 
Funding for the Kepler mission is provided by the NASA Science Mission Directorate. 
STScI is operated by the Association of Universities for Research in Astronomy, Inc., under NASA contract NAS 5–26555. 
This research made use of Lightkurve, a Python package for Kepler and TESS data analysis (Lightkurve Collaboration, 2018). 
\end{ack}

\end{document}